# Balancing Cost and Dissatisfaction in Online EV Charging under Real-time Pricing

(Technical Report)

Hanling Yi[†‡*], Qiulin Lin[†*], Minghua Chen[†]
[†]Department of Information Engineering, The Chinese University of Hong Kong, Hong Kong
[‡]Huawei Noah's Ark Lab, Hong Kong

*Abstract*—We consider an increasingly popular demand-response scenario where a user schedules the flexible electric vehicle (EV) charging load in response to real-time electricity prices. The objective is to minimize the total charging cost with user dissatisfaction taken into account. We focus on the online setting where neither accurate prediction nor distribution of future real-time prices is available to the user when making irrevocable charging decision in each time slot. The emphasis on considering user dissatisfaction and achieving optimal competitive ratio differentiates our work from existing ones and makes our study uniquely challenging. Our key contribution is two simple online algorithms with the best possible competitive ratio among all deterministic algorithms. The optimal competitive ratio is upper-bounded by $\min\left\{\sqrt{\alpha/p_{\min}}, p_{\max}/p_{\min}\right\}$ and the bound is asymptotically tight with respect to $\alpha$, where $p_{\max}$ and $p_{\min}$ are the upper and lower bounds of real-time prices and $\alpha \geq p_{\min}$ captures the consideration of user dissatisfaction. The bounds under small and large values of $\alpha$ suggest the fundamental difference of the problems with and without considering user dissatisfaction. Simulation results based on real-world traces corroborate our theoretical findings and show that the empirical performance of our algorithms can be substantially better than its theoretical worst-case guarantee. Moreover, our algorithms achieve large performance gains as compared to conceivable alternatives. The results also suggest that increasing EV charging rate limit decreases overall cost almost linearly.

## I. Introduction

Integrating renewable energy sources at large scale is a prioritized focus in the recent development of power systems. In the US, renewable electricity accounts for 67% of electricity capacity addition in 2016, and it mounts to 15.6% of the total electricity generation [1]. Globally, in 2016, the installed renewable generation continues to increase and represents 26% of the total electricity generation, exceeding 6,210 terawatt-hours (TWh) [1]. High penetration of renewable generation, however, challenges the conventional operating principle of the power grid. Specifically, renewable generation, such as wind turbine and solar PV, is highly uncertain and intermittent. As renewable generation fluctuates, the conventional approach requires additional *supply-side* flexibility to balance the supply and demand, to ensure power system reliability at all time [2].

Demand response with dynamic pricing is a modern mechanism for providing the needed flexibility, *from the demand side*, to better accommodate renewable generation. The idea is to incentivize customers to adapt the consumption according to the electricity supply conditions, e.g., to shift load from on-peak to off-peak periods, by leveraging real-time electricity prices [3]. Results from real-world trials have been encouraging. In 2015, in the US, the (retail) demand response programs identify 32 GW flexible load, enough to accommodate at least 14% variation in the renewable generation [4] [1]. Furthermore, the rapid and widespread adoption of electric vehicles (EV) introduces substantial flexible charging load that can be leveraged in future demand response programs.[1]

Motivated by the above observations, we focus on an increasingly popular scenario of EV demand response, where individual customers schedule the flexible EV charging demand in response to real-time electricity prices. The objective is to minimize the EV charging cost with user dissatisfaction taken into account.[2] The dissatisfaction captures the user's willingness for the EV not being fully charged, and it is usually encapsulated using the concept of disutility function [6], [7]. With proper real-time pricing in place, minimizing the overall cost of customers (by charging at cheaper time spots) will also benefit the power system because of the reduced peak load and flattened load-curve. Meanwhile, when the prices are relatively high during the entire EV charging period, customers may prefer to charge the EV just enough to meet certain level of satisfaction, so as to avoid excessive electricity cost and casting full charging load to the grid when the supply is scarce or expensive. Overall, it is win-win, for the customers and the system operator, to jointly consider the charging cost and user dissatisfaction.

A number of solutions have been proposed for the offline/stochastic setting where the complete or distributional information of the future real-time prices are known in advance; see a brief summary in Sec. II. While the solutions usually achieve strong performance guarantee, they may not be practical since it is difficult to obtain accurate estimates or distributions of future real-time prices. To illustrate, we plot the real-time prices of two consecutive Tuesdays in the same region in Fig. 1. As seen, the real-time prices in two similar days do not follow similar trajectory patterns or histograms. Thus, it will be challenging to predict real-time prices or their distributions by leveraging the commonly-used information including weekdays and historical price data.

---

[*]The two authors contribute equally to the work.

[1]For example, the annual EV sales in the US have increased by 6x since 2010 and will reach 28% of the US vehicle market by 2031 [5]. The power required for charging a single EV is 2 - 6 KW and can be as high as 20 KW for fast charging models. The charging load is highly flexible and controllable.

[2]With the two-way communication smart grid infrastructure and smart EV charger, the price-aware strategy can be implemented on the vehicle and executed autonomously during the charging.

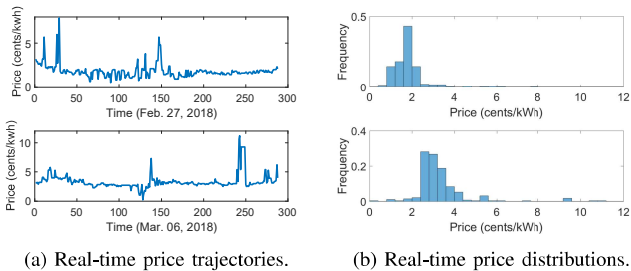

(a) Real-time price trajectories.   (b) Real-time price distributions.

Figure 1: Trajectories and daily distributions of the real-time electricity prices from ComEd in Illinois, USA at 5-minute intervals, for two consecutive Tuesdays Feb 27, 2018 and Mar 06, 2018 with similar weather [8].

In this paper, we investigate the EV charging scheduling problem under a more realistic online setting, where neither accurate prediction nor distribution of future real-time prices is available to users when making online charging decision. Under the online setting, the EV owner has to make irrevocable charging decisions at current time without knowing whether the prices will become cheaper or more expensive in the future. It is challenging as the quality of charging decisions at present time depends on future electricity prices. There have been several pioneering studies along this direction; see a summary in Sec. II. In particular, the authors in [9] propose an online algorithm for optimizing a cost-induced function for EV demand response with performance guarantee that requires little priori knowledge, which makes it amenable for practical implementation.

To this end, we further explore the design of online scheduling algorithms for EV charging under real-time pricing. Our study emphasizes on minimizing charging cost with customer dissatisfaction taken into account and deriving strong/optimal performance guarantees. We employ a new technique to tackle the challenges and design two optimal online algorithms with the best possible competitive ratio among all deterministic ones. We summarize our key contributions as follows.

▷ We formulate the EV charging scheduling problem with both cost and dissatisfaction in consideration in Sec. III. We also consider charging rate limit in the formulation. We then discuss an optimal offline algorithm and the impact of dissatisfaction consideration on the charging behavior and the overall cost in Sec. IV.

▷ In Sec. V, as the key contribution of our study, we design two online algorithms that achieve the best possible competitive ratio among all deterministic algorithms. Along the way, we develop a general technique for designing online optimal algorithms, from an idea hinted in [10]. The optimal competitive ratio is upper-bounded by $\min\{\sqrt{\alpha/p_{\min}}, p_{\max}/p_{\min}\}$ and the bound is asymptotically tight with respect to $\alpha$, where $p_{\max}$ and $p_{\min}$ are the upper and lower bounds of real-time prices and $\alpha \geq p_{\min}$ captures the consideration of user dissatisfaction. Larger (resp. smaller) $\alpha$ indicates more (resp. less) user demand will be satisfied, which represents less (resp. more) user dissatisfaction. The bounds of competitive-ratio give an explicit characterization of the influence of the user dissatisfaction consideration. Furthermore, the bounds under small and large values of $\alpha$ suggest the fundamental difference of the problems with considering user dissatisfaction ($\alpha$ takes small values) and without ($\alpha$ takes large values).

▷ In Sec. VI, we carry out extensive simulation to evaluate the performance of our algorithms based on real-world traces. The results corroborate our theoretical findings and show that the empirical online-to-offline performance ratio of our algorithms is much smaller than competitive ratio. Moreover, our algorithms achieve substantial performance gains as compared to conceivable alternatives. The results also suggest that increasing the EV charging rate limit decreases the overall cost almost linearly.

Due to the space limitation, all proofs are included in our technical report [11].

## II. Related Work

EVs have been widely recognized to have great potential for demand response (DR) [7], [12], [13]. Conventionally, an EV can participate in DR either directly through the time-based pricing scheme set by the system operator (a.k.a. single-level market) [13], or indirectly through the charging station (a.k.a. two-level market) [7]. We focus on the former setting, where the real-time price is served as a signal for DR. A number of solutions have been proposed for DR with EV in the offline/stochastic settings, where the complete or distributional information of the future real-time prices is known in advance [6], [14], [15]. In this paper, we focus on the online setting which assumes minimum knowledge on prices.

Prior to our work, online algorithms have been proposed for EV charging in different scenarios. Majority of the previous work in this area have been focusing on online scheduling of EV charging for charge stations [16]–[18]. For example, Zhao et. al. [16] study competitive online algorithm for aggregator to minimize peak electricity procurement from the grid. Tang et. al. [18] propose an online coordinated charging algorithm for EV charging station to minimize the energy cost. Moreover, there are a few work focusing on the charging strategy design for single EV. Deng et al. [9] study the online cost minimization problem for a single EV charging scenario and propose a near-optimal online algorithm based on a Primal and Dual framework. Later Deng et al. consider both charging and discharging strategies for single EV [19]. We note that our problem is different from theirs in that we take into consideration the trade-off between cost-minimization and user dissatisfaction. In addition, our algorithms can achieve the optimal competitive ratio among all deterministic online algorithms.

## III. EV Charging Problem Formulation

We consider the EV charging problem within a charging period $T$. The charging period is divided into slots with equal lengths, denoted by $[T] \triangleq \{1, 2, ..., T\}^3$. Due to real-time electricity pricing, the charging price of EV is time varying. At slot $t$, when the charging price is $p(t)$ and the charging quantity (in $kWh$) is $v(t)$, the EV incurs a charging cost of $p(t)v(t)$. For ease of presentation, we assume $v(t) \in [0, 1]$, i.e., $v(t)$ is the charging quantity normalized by the maximum charging quantity in a single slot[4]. Denote $c > 0$ as the battery capacity normalized by the maximum charging quantity within a single slot. If the battery is fully charged at the end of the charging period, i.e., $\sum_{t=1}^{T} v(t) = c$, the EV owner incurs total charging cost of $\sum_{t=1}^{T} p(t)v(t)$. Otherwise, the EV owner suffers an additional penalty, named user dissatisfaction, which is modeled as a linear function[5] on the uncharged capacity $c - \sum_{t=1}^{T} v(t)$. In summary, the EV charging problem, denoted as $EVC$, can be formulated as the following:

$$EVC: \quad \min \sum_{t=1}^{T} p(t)v(t) + \alpha \left( c - \sum_{t=1}^{T} v(t) \right)$$

$$s.t. \sum_{t=1}^{T} v(t) \leq c, \quad (1)$$

$$0 \leq v(t) \leq 1, \forall t \in [T] \quad (2)$$

The constraint in (1) indicates that the total charging quantity can not exceed the battery capacity. The parameter $\alpha \geq p_{\min}$ captures the consideration of user dissatisfaction[6]. The objective function in $EVC$ consists of two terms: the first is the charging cost, and the second represents the user dissatisfaction. We call the objective value of $EVC$ as *cost-plus-dissatisfaction value*. Both $c$ and $\alpha$ are specified by the EV owner in advance.

The offline version of $EVC$ is a linear problem (LP), which can be solved efficiently. However, in practice, the problem inputs such as the charging period and the real-time electricity prices are not known in advance. Thus we are interested in the online setting, i.e., $T$ and $\sigma^{[T]} \triangleq (p(1), p(2), ..., p(T)) \in \Sigma^{[T]}$ are revealed sequentially, where $\Sigma^{[T]}$ is the set of all possible price sequence in $T$ slots. We assume the real-time electricity prices are bounded, i.e., $p(\tau) \in [p_{\min}, p_{\max}], \forall \tau \in [T]$. Here $p_{\min}, p_{\max}$ are common knowledge and we denote $\theta \triangleq p_{\max}/p_{\min}$. In this paper, we aim to develop deterministic online algorithms that have the optimal competitive ratio which is defined as the best achievable online to offline performance ratio in the worst case. Formally, the competitive ratio of an online algorithm $\mathcal{A}$ can be defined as

$$CR_{\mathcal{A}} = \max_{\sigma^{[T]} \in \Sigma^{[T]}} \frac{\eta_{\mathcal{A}}}{\eta_{OPT}}, \quad (3)$$

where $\eta_{OPT}$ and $\eta_{\mathcal{A}}$ are the cost-plus-dissatisfaction values for offline and online algorithm under input $\sigma^{[T]}$, respectively. The best competitive ratio is then defined as $CR = \min_{\mathcal{A}} CR_{\mathcal{A}}$.

**Remark:** Note that in our formulation, we assume the real-time price is not affected by the charging quantity of EV. This is a common assumption in demand response literature [3], [6], [13]. On one hand, it is reasonable as we are considering the single EV charging problem and the charging quantity of a single EV has negligible impact on the power system supply-and-demand balance. The study to investigate the aggregate impact of a large number of EVs on the real-time price would be an interesting future work. On the other hand, a complex model on the real-time price would require the EV owner to constantly update the model parameters, as market condition may change from time to time. This is not practical in the resource-limited scenario, which is the case for most EV owners.

## IV. OPTIMAL OFFLINE ALGORITHM

In this section, we consider the problem under the offline setting where the prices are given in advance. We discuss the optimal offline algorithm, and the impact of dissatisfaction consideration on the charging behavior and the cost-plus-dissatisfaction value.

Let us first consider the offline algorithm for the EV charging problem without charging rate limit constraint, i.e., we omit $v(t) \leq 1$ in (2). Denote the input until time $t$ as $\sigma^{[t]} \triangleq (p(1), p(2), ..., p(t)) \in \Sigma^{[t]}$, and the minimum price observed in the first $t$ slots as $p_{\min}^t \triangleq \min_{\tau \in [t]} p(\tau)$. Given $\sigma^{[t]}$, the offline problem at time $t$ is an LP. We know that when $\alpha \leq p_{\min}^t$, it is optimal to not charge the EV; otherwise, it is best to fully charge the EV at price $p_{\min}^t$. Thus the optimal offline cost-plus-dissatisfaction value until time $t$ is given by

$$\widecheck{OPT}(t) = \min\left\{ p_{\min}^t, \alpha \right\} \cdot c. \quad (4)$$

From the above analysis, we can see the impact of $\alpha$ on the offline solution. When taking into account the user dissatisfaction in consideration, sometimes it is better not to fully charge the EV.

Secondly, let us study the offline algorithm for problem $EVC$ (i.e., with charging rate limit $v(t) \leq 1, \forall t \in [T]$). For ease of presentation, we consider the case when $c$ is a positive integer. Due to the charging rate limit, it may take multiple slots to fully charge a EV and we prefer to charge the EV in the slots with low prices. Formally, at time $t$, if $t \geq c$, we denote the set $\mathcal{T}_t$ as the index set of the $c$-smallest prices in $\sigma^{[t]}$ (break ties randomly); otherwise we denote $\mathcal{T}_t = [t]$. Clearly, we have $|\mathcal{T}_t| \leq c$. Further, define the set $\mathcal{T}_t^\alpha \triangleq \{\tau | p(\tau) < \alpha, \tau \in [t]\}$, i.e., $\mathcal{T}_t^\alpha$ is the set of slots in which the prices are smaller than $\alpha$. Then at slot $t$, the optimal offline solution under $\sigma^{[t]}$ is given as follow: $\forall \tau \in [t]$,

$$v^*(\tau) = \begin{cases} 1, & \tau \in \mathcal{T}_t \cap \mathcal{T}_t^\alpha; \\ 0, & \text{otherwise.} \end{cases} \quad (5)$$

Therefore, the optimal cost-plus-dissatisfaction value at time $t$, denoted as $OPT(t)$, can be expressed as

$$OPT(t) = \sum_{\tau \in \mathcal{T}_t \cap \mathcal{T}_t^\alpha} p(\tau) + \alpha \left( c - \sum_{\tau \in \mathcal{T}_t \cap \mathcal{T}_t^\alpha} 1 \right). \quad (6)$$

Note that when $|\mathcal{T}_t \cap \mathcal{T}_t^\alpha| = c$, the expression of $OPT(t)$ can be simplified as $OPT(t) = \sum_{\tau \in \mathcal{T}_t \cap \mathcal{T}_t^\alpha} p(\tau)$. From (5), we know the offline solution depends on $\alpha$. If $\alpha$ is large, for example, $\alpha \geq p_{\max}$, then $\mathcal{T}_t^\alpha = [t]$ and the EV will be charged in slots that belong to $\mathcal{T}_t$. If $\alpha$ is small, it is possible that the EV is not fully charged at the end. The rationality behinds this is that the EV owner prefers to charge the EV just enough to meet certain level of satisfaction when the prices are relatively high (as compare to $\alpha$) to save the charging cost. In fact, the EV will be fully charged at time $t$ only when there are no less than $c$ slots at which the prices are lower than $\alpha$. In addition, the system operator also benefits as the demand is reduced during periods with high prices indicating scare or expensive supply. This leads to a win-win situation.

---

[3] In this paper, we use the notation $[n], n \in \mathbb{Z}^+$ to denote the set $\{1, 2, ..., n\}$.

[4] The maximum charging quantity in single slot is the battery charging rate limit multiplies by the time duration in a single slot, i.e., 5 minutes.

[5] A linear dissatisfaction model mimics the charging cost model. We leave considering a more general model as our future work.

[6] Note that when $\alpha < p_{\min}$, the optimal solution is to never charge the EV, which is trivial. Therefore, we only consider $\alpha \geq p_{\min}$.

## V. Optimal Online Algorithms

In this section, we study online algorithms for *EVC*. We firstly study the problem when there is no charging rate limit and propose two optimal online algorithms in Sec. V-A. We then extend these two online algorithms to the case with charging rate limit by employing a neat divide-and-conquer idea in Sec. V-B and showing that they achieve the best possible competitive ratio among all deterministic online algorithms for *EVC*.

### A. Without Charging Rate Limit

For ease of presentation, we denote the EV charging problem without charging rate limit as $EVC_{NC}$. We develop a general technique for online algorithm design and propose two optimal deterministic online algorithms for $EVC_{NC}$. The two proposed online algorithms have the same online-to-offline performance ratio in the worst-case. Meanwhile, the first one is too conservative under general cases, thus we propose the second online algorithm as an improved version of the first one and show that it has better performance in general.

*1) Optimal Online Algorithm for $EVC_{NC}$:* Given any $\pi \geq 1$, we define a class of online algorithm $ALG(\pi)$ as follows: at time $t$, the output of $ALG(\pi)$ is given by

$$v(t) = \frac{[\eta^{t-1} - \tilde{OPT}(t)\pi]^+}{\alpha - p(t)}, \quad (7)$$

where $[x]^+ = \max\{0, x\}$, $\tilde{OPT}(t)$ is the cost-plus-dissatisfaction value of offline algorithm defined in (4), and $\eta^{t-1}$ is the cost-plus-dissatisfaction value of online algorithm up to time $t-1$. Clearly we have $\eta^0 = \alpha c$ and

$$\eta^t = \eta^{t-1} - (\alpha - p(t))v(t). \quad (8)$$

From (7) and (8), we know that $ALG(\pi)$ can guarantee at each time $t$, the online-to-offline performance ratio is bounded by $\pi$, i.e., $\eta^t/\tilde{OPT}(t) \leq \pi$. In particular, when $\eta^{t-1}/\tilde{OPT}(t) \leq \pi$, $ALG(\pi)$ does not charge the EV, i.e., $v(t) = 0$. In this case, we have $\eta^t = \eta^{t-1}$ and thus $\eta^t/\tilde{OPT}(t) \leq \pi$. When $\eta^{t-1}/\tilde{OPT}(t) > \pi$, $ALG(\pi)$ will charge the EV[7], i.e., $v(t) = \frac{\eta^{t-1} - \tilde{OPT}(t)\pi}{\alpha - p(t)}$ and in this case we have $\eta^t/\tilde{OPT}(t) = \pi$. Thus intuitively, the online algorithm $ALG(\pi)$ will charge the EV only when the ratio $\eta^{t-1}/\tilde{OPT}(t)$ is larger than $\pi$. In that case, $ALG(\pi)$ charges just enough at each time to keep the ratio to be $\pi$. The idea of keeping the online-to-offline performance ratio in $ALG(\pi)$ originates from the discussion in the two intuitive rules of designing online algorithms for the one-way trading problem [10]. In the following, we show that this simple online algorithm $ALG(\pi)$ with a carefully chosen $\pi^*$ is optimal among all the deterministic online algorithms for $EVC_{NC}$.

---

[7]In this case we must have $p(t) < \alpha$, otherwise according to (4), the optimal offline cost-plus-dissatisfaction value remains unchanged, then we have $\tilde{OPT}(t) = \tilde{OPT}(t-1) \geq \eta^{t-1}/\pi$ is a contradiction.

**Definition 1.** For $ALG(\pi)$ with $\pi \geq 1$. If for any $T$ and $\sigma^{[T]}$, we have $\sum_{t=1}^{T} v(t) \leq c$, then we say $ALG(\pi)$ is feasible.

We note that there always exists a $\pi$ (large enough) that can guarantee the feasibility of $ALG(\pi)$. For example, when $\pi \geq \frac{\alpha}{p_{\min}}$, $\eta^0 = \alpha c \leq \pi p_{\min} c \leq \pi \tilde{OPT}(t), \forall t$ and any input, which means $ALG(\pi)$ can maintain a competitive ratio $\pi$. As such, we easily conclude $v(t) = 0, \forall t \in [T], \sigma^{[T]}$ and thus $ALG(\pi)$ is feasible. For any $\pi \geq 1$, we can define the maximum total charging quantity for $ALG(\pi)$ as

$$V(\pi) \triangleq \max_{\sigma^{[T]} \in \Sigma^{[T]}} \sum_{\tau=1}^{T} v(\tau), \quad (9)$$

where $v(\tau), \forall \tau \in [T]$ are the outputs of $ALG(\pi)$ defined in (7). By definition, $ALG(\pi)$ is feasible if $V(\pi) \leq c$. It is obvious that if $ALG(\pi)$ is feasible, then it is $\pi$-competitive. Denote the set $\mathcal{F} \triangleq \{\pi | V(\pi) \leq c\}$, i.e., $\mathcal{F}$ is the set of $\pi$ such that $ALG(\pi)$ is feasible. In the following, we are dedicated to find the minimum $\pi^*$ such that $ALG(\pi^*)$ is feasible. Namely, we want to find $\pi^* = \min_{\pi \in \mathcal{F}} \pi$.

**Lemma 2.** *To compute $V(\pi)$, it is sufficient to consider decreasing price sequences with the first price $p_1$ satisfies $p_1 \leq \min\left\{\frac{\alpha}{\pi}, p_{\max}\right\}$.*

Intuitively, following $ALG(\pi)$, when $p(t) \geq p(t-1)$, $\tilde{OPT}(t) = \tilde{OPT}(t-1)$, i.e., the optimal offline cost-plus-dissatisfaction remains unchanged and thus $ALG(\pi)$ will not charge at slot $t$ to keep $\pi$. As a result, we can delete such $p(t)$ from the price sequence and the total charging quantity won't change. For ease of presentation, in the following, we consider an arbitrary decreasing price sequence as the following:

$$\min\left\{\frac{\alpha}{\pi}, p_{\max}\right\} \geq p(1) > p(2) > \cdots > p(T) \geq p_{\min}. \quad (10)$$

We denote the decreasing price sequence in (10) as $\bar{\sigma}^{[T]} = (p(1), p(2), ..., p(T))$.

**Lemma 3.** *For any $\pi > 0$, we have*

$$V(\pi) = \begin{cases} 0 & , \frac{\alpha}{\pi} \leq p_{\min}; \\ c\pi \ln \frac{\alpha - p_{\min}}{\alpha - \frac{\alpha}{\pi}} & , p_{\min} \leq \frac{\alpha}{\pi} \leq p_{\max}; \\ \frac{\alpha c - p_{\max} c\pi}{\alpha - p_{\max}} + c\pi \ln \frac{\alpha - p_{\min}}{\alpha - p_{\max}} & , \frac{\alpha}{\pi} > p_{\max}. \end{cases} \quad (11)$$

*Moreover, $V(\pi)$ is decreasing in $\pi$.*

Following Lemma 3, we know that the minimum $\pi^*$ should satisfy $V(\pi^*) = c$. Since $V(1) = c + c \ln \frac{\alpha - p_{\min}}{\alpha - p_{\max}} > c$ and $\lim_{\pi \to +\infty} V(\pi) < c$, from Lemma 3, we know that $\pi^*$ is unique. We plot function $V(\pi)$ in Fig. 2, under the setting specified in the caption. From this figure, we can observe that $V(\pi)$ is decreasing in $\pi$. By solving for the equation $V(\pi^*) = c$, we can get the expression of $\pi^*$, as summarized in the following theorem.

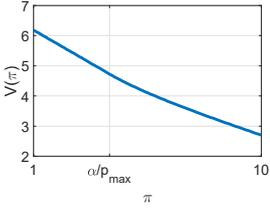
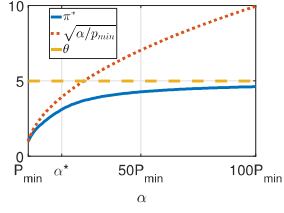

Figure 2: The plot of $V(\pi)$ for the case where $p_{\min} = 1$, $\theta = 5$, $\alpha = 4p_{\max}$ and $c = 5$.

Figure 3: $\pi^*$ and the upper bound under the setting where $p_{\min} = 1$ and $\theta = 5$.

**Theorem 4.** *Denote $\alpha^*$ as the root for equation $\frac{\alpha}{p_{\max}} \ln \frac{\alpha - p_{\min}}{\alpha - p_{\max}} = 1$, which exists and is unique. Then when $\alpha > \alpha^*$, we have*

$$\pi^* = \frac{\frac{p_{\max}}{\alpha - p_{\max}}}{\frac{p_{\max}}{\alpha - p_{\max}} - \ln \frac{\alpha - p_{\min}}{\alpha - p_{\max}}}. \tag{12}$$

*Otherwise, $\pi^*$ is the unique root for the following equation*

$$\pi^* \ln \frac{\alpha - p_{\min}}{\alpha - \frac{\alpha}{\pi^*}} = 1. \tag{13}$$

*Further, $\pi^*$ is upper-bounded by $\min\left\{\sqrt{\alpha/p_{\min}}, \theta\right\}$ and the bound is asymptotically tight with respect to $\alpha$.*

Recall that $\theta = p_{\max}/p_{\min}$. To further illustrate the relationship between $\alpha$ and $\pi^*$, in Fig. 3 we plot $\pi^*$ and its upper bound $\sqrt{\alpha/p_{\min}}$. Clearly, we can observe that $\pi^* \leq \min\left\{\sqrt{\alpha/p_{\min}}, \theta\right\}$ and it approaches $\theta(= 5)$ as $\alpha$ increases. The competitive-ratio bound gives an explicit characterization of the influence of the user dissatisfaction consideration. Specifically, when $\alpha$ is small, i.e., $\alpha \in [p_{\min}, p_{\max}\theta]$, $\pi^*$ is upper bounded by $\sqrt{\alpha/p_{\min}}$. When $\alpha$ is large, i.e., $\alpha \in (p_{\max}\theta, +\infty)$, $\pi^*$ is upper bounded by $\theta$. In particular, when $\alpha$ approaches infinity, the optimal competitive ratio approaches $\theta$. This contrasting expressions of the competitive-ratio bound for small and large $\alpha$ suggest a fundamental difference between the problems with and without user dissatisfaction in consideration.

**Theorem 5.** *$ALG(\pi^*)$ achieves the optimal competitive ratio among all the deterministic online algorithms.*

We prove theorem 5 by presenting the worst-case input of $ALG(\pi^*)$ to any deterministic online algorithm $\mathcal{A}$, and show that under this input, the competitive ratio of $\mathcal{A}$ is at least $\pi^*$.

*2) Adaptive Online Algorithm for $EVC_{NC}$:* In this subsection, we propose an adaptive online algorithm $ALG$ that has the same competitive ratio as $ALG(\pi^*)$. Meanwhile, it improves the online-to-offline performance ratio under general-case inputs.

From (7), we know that $ALG(\pi^*)$ charges just enough energy at each slot to keep the online-to-offline performance ratio to be no larger than $\pi^*$. Following $ALG(\pi^*)$, it is guaranteed that the EV will be fully charged with the real-time price under the worst case input (the input such that $V(\pi^*) = c$). However, in practice, it is more likely that we will encounter the non-worst case input. Under a general-case input, the EV is not fully charged.

Consider the following motivating example, when $\alpha = p_{\max}$ and the initial price is $p_{\min}$, from (13) we know $\pi^* > 1$. According to (7), the output of $ALG(\pi^*)$ is

$$v(1) = \frac{\alpha c - p_{\min} c \pi^*}{\alpha - p_{\min}} < c.$$

The cost-plus-dissatisfaction of $ALG(\pi^*)$ is $\eta^1 = p_{\min} c \pi^*$. However, $ALG(\pi^*)$ would have achieved a lower cost-plus-dissatisfaction if it fully charges the EV at price $p_{\min}$.

Intuitively, a "good" online algorithm will try to charge the EV more when the price is low, i.e., being more opportunistic. However, by design $ALG(\pi^*)$ is pessimistic: it only tries to maintain the competitive ratio to be no larger than $\pi^*$, even if for some (general-case) inputs one can certainly do better than $\pi^*$. Thus one way to improve the non-worst case performance of $ALG(\pi^*)$ is as follows: instead of trying to keep the online-to-offline performance ratio as $\pi^*$ during the whole charging period, at time $t$, the online algorithm $ALG$ chooses a $\pi_t$ to maintain. This $\pi_t$ is chosen as the smallest attainable competitive ratio at time $t$, given the previous inputs $\sigma^{[t]}$ and outputs $v(\tau), \forall \tau \in [t-1]$ of the algorithm, and taking into account the possible inputs in future slots. As compared to $ALG(\pi^*)$, $ALG$ can adaptively change $\pi^*$ at each slot. Obviously, under worst case input, $\pi_t = \pi^*, \forall t \in [T]$, i.e., $ALG$ performs exactly the same as $ALG(\pi^*)$. Thus $ALG$ has the same competitive ratio as $ALG(\pi^*)$. However, under non-worst case input, we have $\pi_t < \pi^*$, i.e., $ALG$ strictly improves the online-to-offline performance ratio.

We now present the adaptive algorithm $ALG$ and study its performance. At time $t$, $ALG$ computes the best attainable competitive ratio $\pi_t^*$. Given $\pi_t^*$, if $p(t) \geq \alpha$, the output of $ALG$ is 0. Otherwise, the output of $ALG$ is

$$v(t) = \frac{[\eta^{t-1} - \tilde{OPT}(t)\pi_t^*]^+}{\alpha - p(t)}, \tag{14}$$

where $\eta^{t-1}$ is the online cost-plus-dissatisfaction value of $ALG$ up to time $t-1$. From (14), we can observe that the structure of $ALG$ and $ALG(\pi^*)$ are the same, except that in $ALG$, the target ratio $\pi_t^*$ is changing with $t$. The question remained is how to determine the best attainable competitive ratio $\pi_t^*$? In the following, we are dedicated to find $\pi_t^*$ and study its property.

Similarly from Lemma 2, we know it is sufficient to consider decreasing price sequences. In the following, we consider the decreasing price sequence $\bar{\sigma}^{[T]}$ defined in (10). At time $t$, we denote

$$V_t(\pi_t) \triangleq \max_{\bar{\sigma}^{[t+1:T]} \in \Sigma^{[t+1:T]}} \sum_{\tau=1}^{T} v(\tau) = \sum_{\tau=1}^{t} v(\tau) + \max_{\bar{\sigma}^{[t+1:T]} \in \Sigma^{[t+1:T]}} \sum_{\tau=t+1}^{T} v(\tau).$$

Note that in $V_t(\pi_t)$ we are optimizing over $\bar{\sigma}^{[t+1:T]}$ since the input $\bar{\sigma}^{[t]}$ and the output $v(\tau), \forall \tau \in [1, t-1]$ are given at time $t$. By definition, we know $V_0(\pi_0) = V(\pi)$, thus it is easy to conclude that $\pi_0^* = \pi^*$.

**Lemma 6.** *Under input $\bar{\sigma}^{[T]}$, at time $t \in [T]$, for any given $\pi_t$, we have*

$$V_t(\pi_t) = \sum_{\tau=1}^{t-1} v(\tau) + \frac{\eta^{t-1} - p(t)c\pi_t}{\alpha - p(t)} + c\pi_t \ln \frac{\alpha - p_{\min}}{\alpha - p(t)} \tag{15}$$

and $V_t(\pi_t)$ is decreasing in $\pi_t$.

From Lemma 6, we know that at time $t$, the optimal $\pi_t^*$ should satisfy $V_t(\pi_t^*) = c$. Then we can obtain the expression for $\pi_t^*$ by solving the equation $V_t(\pi_t^*) = c$, as shown in the following theorem.

**Theorem 7.** *At time $t$, we have*

$$\pi_t^* = \frac{c - \sum_{\tau=1}^{t-1} v(\tau) - \frac{\eta^{t-1}}{\alpha - p(t)}}{c \ln \frac{\alpha - p_{\min}}{\alpha - p(t)} - \frac{cp(t)}{\alpha - p(t)}}, \quad (16)$$

*and $\pi_t^*$ is non-increasing in $t$.*

Note that $\pi_t^*$ is a function of $v(\tau), \forall \tau \in [t-1]$ and $p(\tau), \forall \tau \in [t]$. It is not hard to observe that $\pi_1^* \le \pi^*$ and $\pi_1^* = \pi^*$ only when the initial price is $\min\left\{\frac{\alpha}{\pi^*}, p_{\max}\right\}$. Thus from theorem 7, we know that $ALG$ performs better than $ALG(\pi^*)$ under general-case inputs.

> Back to our motivating example, under the same input, from (16), we know that $\pi_1^* = 1$. Thus the output of $ALG$ is $c$, i.e., $ALG$ fully charges the EV at price $p_{\min}$ and achieves an online-to-offline performance ratio of 1.

We end the discussion in this subsection by summarizing the general technique we used for designing online algorithms, which can be of independent interest: at each slot, the online algorithm will try to maintain the online-to-offline performance ratio to be no larger than a target value. The target value is chosen as the minimum possible value that the online algorithm can maintain under all the possible uncertain inputs, and it can either be fixed (as in $ALG(\pi^*)$) or adaptively changed (as in $ALG$). The competitive ratio of the online algorithm is by definition the target value under worst case input (for example, the competitive ratio of $ALG$ is $\max_{\sigma^{[T]} \in \Sigma^{[T]}} \pi_T$). In our recent work [20], we develop a framework named CR-Pursuit for solving online revenue maximization problems with inventory constraint based on similar idea.

### B. With Charging Rate Limit

In this subsection, adapting the results from Sec. V-A, we propose two optimal online algorithms for $EVC$ based on a divide-and-conquer technique. We note that the extension from non-charging-rate-limit case to with-charging-rate-limit case is nontrivial, and many related works [15], [16] do not consider charging rate limit when designing EV charging algorithms.

First, we observe that when $c \le 1$, problem $EVC$ is equivalent to $EVC_{NC}$. To see this, note that when $c \le 1$, the constraint (1) in $EVC_{NC}$ becomes $\sum_{t=1}^{T} v(t) \le c \le 1$, which together with $v(t) \ge 0$, implies that $v(t) \le 1$. With this observation, from Sec. V-A, we have the following corollary:

**Corollary 8.** *$ALG(\pi^*)$ and $ALG$ achieve the optimal competitive ratio among all deterministic online algorithms for $EVC$ when $c \le 1$.*

We then utilize this observation to design optimal online algorithms for $EVC$ when $c$ is a positive integer. Later we will show that our analysis can be easily extended to the case when $c$ is a positive rational number. Further, when $c$ is a positive real number, we can run our algorithm with the rational number $c'$ that is arbitrarily close to $c$, and it can be shown that the competitive ratio is arbitrarily close to the optimal competitive ratio.

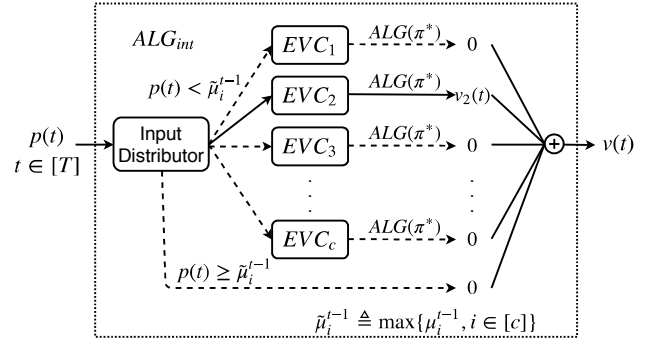

Figure 4: Schema of algorithm $ALG_{int}$. At each slot, the real-time price is assigned to one sub-problem. For a particular moment, the price $p(t)$ is assigned to $EVC_2$.

*1) c is a Positive Integer:* We now introduce our online algorithm, denoted as $ALG_{int}$, for the case when $c$ is a positive integer. The main idea behind $ALG_{int}$ is divide and conquer. Specifically, we decompose problem $EVC$ into $c$ sub-problems, denoted as $EVC_i, i \in [c]$, where $[c] \triangleq \{1, 2, \ldots, c\}$. Each sub-problem is a problem $EVC$ with $c = 1$ and its design variable at time $t$ is denoted as $v_i(t)$. The real-time prices are distributed to the sub-problems according to an *input distributor*. For each sub-problem, we runs the online algorithm $ALG(\pi^*)$[8], which outputs $v_i(t)$, if assigned with an input at slot $t$ from the input distributor; 0, otherwise. The outputs to the sub-problems are then combined to attain a solution to the original problem.

We now discuss how the input distributor works in $ALG_{int}$. Let $\mu_i^t$ be the latest price distributed to $EVC_i$ up to slot $t$ (inclusive). Initially, we set $\mu_i^0 = \alpha, \forall i \in [c]$. At slot $t$, the input distributor in $ALG_{int}$ works as follows: when the real-time price $p(t)$ is revealed, compare $p(t)$ with $\tilde{\mu}_i^{t-1} = \max\{\mu_i^{t-1}, i \in [c]\}$. If $p(t) \ge \tilde{\mu}_i^{t-1}$, then $p(t)$ is discarded and the output at time $t$ is 0. Otherwise, the input distributor assigns $p(t)$ to sub-problem $EVC_{i^*}$, where $i^* = \arg\max_{i \in [c]} \mu_i^{t-1}$ (we select the minimum one if there are multiple solutions) and we must have $p(t) < \mu_{i^*}^{t-1}$. Note that at slot $t$, $\mu_{i^*}^t = p(t)$ and for other sub-problems ($i \ne i^*$), $\mu_i^t = \mu_i^{t-1}$.

For $EVC_i$, denote the output at time $t \in [T]$ as $v_i(t)$, the corresponding online cost-plus-dissatisfaction value as $\eta_i^t$ and offline optimal cost-plus-dissatisfaction value as $OPT_i(t)$. If $EVC_i$ is not assigned a price at slot $t$, it does nothing (i.e., $ALG(\pi^*)$ skips this slot). In this case, we specify $v_i(t) = 0$; consequently, $\eta_i^t = \eta_i^{t-1}$ and $OPT_i(t) = OPT_i(t-1)$. Otherwise, $EVC_i$ is assigned a price $p(t)$. In this case, $p(t) \le \mu_i^{t-1}$ (latest price distributed to $EVC_i$ up to slot $t - 1$), which holds for all the slots that $EVC_i$ is assigned a price, and thus we easily conclude the prices assigned to $EVC_i$ are decreasing in time.

---

[8]Or alternatively, $ALG$. The analysis when $ALG_{int}$ use $ALG(\pi^*)$ or $ALG$ is the same. In the following, we only focus on the analysis when $ALG_{int}$ uses $ALG(\pi^*)$.

**Algorithms 1** Online Algorithm $ALG_{int}$ for $EVC$.

1: $\mu_i^0 = \alpha, \forall i \in [c]$;
2: At slot $t$, $p(t)$ is revealed;
3: Set $\mu_i^t = \mu_i^{t-1}, \forall i$
4: **if** $p(t) \geq \max\{\mu_i^{t-1}, i \in [c]\}$ **then**
5:     $v(t) = 0$
6: **else**
7:     Assign $p(t)$ to $EVC_{i^*}$ ($i^* = \arg\max_{i \in [c]} \mu_i^{t-1}$);
8:     Set $\mu_{i^*}^t = p(t)$;
9:     Attain output $v_{i^*}(t)$ from $EVC_{i^*}$ by Alg. 2;
10:    Set $v(t) = v_{i^*}(t)$;
11: **end if**

---

**Algorithms 2** Online Algorithm $ALG(\pi^*)$ for $EVC_i$.

1: $\eta_i^0 = \alpha$, $OPT_i(0) = \alpha$;
2: At slot $t$,
3: **if** Assigned $p(t) < \alpha$ **then**
4:     $OPT_i(t) = p(t) = \mu_i^t$;
5:     $v_i(t) = \frac{[\eta_i^{t-1} - OPT_i(t)\pi^*]^+}{\alpha - p(t)}$;
6:     $\eta_i^t = \eta_i^{t-1} - (\alpha - p(t))v_i(t)$;
7: **else**
8:     $v_i(t) = 0$; $OPT_i(t) = OPT_i(t-1)$; $\eta_i^t = \eta_i^{t-1}$;
9: **end if**

---

**Algorithms 3** Online Algorithm $ALG_{rat}$ for $EVC$.

1: $\mu_i^0 = \alpha, \forall i \in [m]$;
2: At slot $t$, $p(t)$ is revealed;
3: Set $\mu_i^t = \mu_i^{t-1}, \forall i \in [m]$
4: **if** $p(t) \geq \max\{\mu_i^{t-1}, i \in [m]\}$ **then**
5:     $v(t) = 0$
6: **else**
7:     $A = \{i, \mu_i^{t-1} > p(t)\}$; $B$: index set of the sub-problem with $n$ largest $\mu_i^{t-1}$ (break ties randomly)
8:     Assign $p(t)$ to $EVC_i$, set $\mu_i^t = p(t)$, $\forall i \in A \cap B$;
9:     Attains output $v_i(t)$ from $EVC_i, \forall i \in [m]$ by Alg. 2;
10:    $v(t) \leftarrow \sum_{i \in [m]} v_i(t)$;
11: **end if**

---

As such, $OPT_i(t) = p(t) = \mu_i^t$. Then, $v_i(t)$ and $\eta_i^t$ can be computed following $ALG(\pi^*)$. An example is illustrated in Fig. 4: at a particular time $t$, $p(t)$ is distributed to $EVC_2$ according to the input distributor. The output of $EVC_2$ is $v_2(t)$, while for the other sub-problems, their outputs are 0. The outputs of all the sub-problems are then combined to produce $v(t)$.

We summarize $ALG_{int}$ for $EVC$ and $ALG(\pi^*)$ for $EVC_i$ in Alg. 1 and Alg. 2, respectively. In the following, we characterize the competitive ratio of online algorithm $ALG_{int}$.

**Lemma 9.** *Following $ALG_{int}$, the online-to-offline performance ratio for each sub-problem $EVC_i$ at any time $t$ for any real-time price sequence is upper bounded by $\pi^*$. That is for any $\sigma^{[1:T]}$, $\eta_i^t \leq \pi^* OPT_i(t), \forall i \in [c], \forall t \in [T]$.*

We then characterize the relationship between the online and offline cost-plus-dissatisfaction values for $EVC$ and $EVC_i, \forall i \in [c]$ in Lemma 10 and 11, respectively.

**Lemma 10.** *At any time $t$, the online cost-plus-dissatisfaction value of $ALG_{int}$ for $EVC$, denoted as $\eta^t$, equals the summation of those of $ALG(\pi^*)$ for each sub-problem, i.e., $\eta^t = \sum_{i=1}^c \eta_i^t, \forall t \in [T]$.*

**Lemma 11.** *Following the input distributor in $ALG_{int}$, at any time $t$, the offline cost-plus-dissatisfaction value for $EVC$ equals the summation of those for each sub-problem, i.e., $OPT(t) = \sum_{i=1}^c OPT_i(t), \forall t \in [T]$.*

The key insights from Lemma 10 and 11 are that (i) our divide-and-conquer approach incurs no optimality loss under both online and offline settings, and (ii) the competitive ratio for the overall problem should also be the same as the individual sub-problem. These insights lead to the following result.

**Theorem 12.** *$ALG_{int}$ is $\pi^*$-competitive.*

*Proof:* By Lemma 9, 10, 11, we have for any input $\sigma^{[1:T]}$, at any $t \in [T]$

$$\eta^t = \sum_{i=1}^c \eta_i^t \leq \pi^* \sum_{i=1}^c OPT_i(t) = \pi^* OPT(t).$$

We can then easily conclude $ALG_{int}$ is $\pi^*$-competitive. ∎

Using similar technique in the proof of the optimality of $ALG(\pi^*)$ in Sec. V-A, we can show that $ALG_{int}$ is the optimal deterministic online algorithm.

**Theorem 13.** *$ALG_{int}$ achieves the optimal competitive ratio among all deterministic online algorithms.*

*2) Extension: c is a Positive Rational Number:* In this subsection, we extend $ALG_{int}$ to the case when $c$ is a positive rational number. Suppose $c = m/n$. If $m \leq n$, it is equivalent to the case without charging rate limit. We thus focus on the case when $m > n$. We again apply the divide-and-conquer approach. First, we define $m$ sub-problems, denoted as $EVC_i$, $i \in [m]$. Each sub-problem is a problem $EVC$ with $c = 1/n$. Each sub-problem also runs Alg. 2. The input distributor works a little differently. Specifically, when a price $p(t)$ is revealed at slot $t$, we compare $p(t)$ with $\{\mu_i^{t-1}, i \in [m]\}$. If $p(t) \geq \max\{\mu_i^{t-1}, i \in [m]\}$, then $p(t)$ is discarded and $v(t)$ is set to be 0. Otherwise, let $A = \{i, \mu_i^{t-1} > p(t)\}$, which represents the index set of sub-problems such that $\mu_i^{t-1} > p(t)$. And let $B$ be the index set of the sub-problems with $n$ largest $\mu_i^{t-1}$ (break ties randomly). Then $p(t)$ is distributed to $EVC_i, \forall i \in A \cap B$. Namely, we distribute $p(t)$ to at most $n$ sub-problems with higher $\mu_i^{t-1}$ (compared with $p(t)$). We set $v(t) = \sum_{i \in [m]} v_i(t)$. This is to make sure the online (resp. optimal offline) cost-plus-dissatisfaction value at time $t$ is always the same as the sum of online (resp. optimal offline) cost-plus-dissatisfaction value of all the sub-problems. The algorithm for $EVC$ when $c = m/n$ ($m \geq n$), denoted as $ALG_{rat}$, is summarized in Alg. 3.

**Theorem 14.** *$ALG_{rat}$ is $\pi^*$-competitive and it is optimal among all deterministic online algorithms.*

## VI. Performance Evaluation

### A. Evaluation Setup

We consider a residential user with an EV, who has enrolled in the real-time pricing (RTP) scheme. The real-time prices are updated every 5 minutes and thus the length of each time slot is 5 minutes in our simulation. We obtain the 5-min real-time prices (from June, 2017 to June, 2018) from ComEd in Illinois, USA [21]. In practice, the real-time prices can be negative, as we can observed from the data. However, we note that these abnormal prices seldom occur, for example, the percentage of negative prices within a year is less than 3%. In our simulation, we only focus on the normal-price setting and we preprocessed the real-time price data by eliminating the largest 5% and smallest 5% prices within the year. After preprocessing, we have $p_{\min} = 1.3$ (cents/kWh) and $\theta = 4.54$ for the one-year price data. We consider a scenario where the EV owner park the EV from 5pm to 8am. The EV has a charging power of 8.8kW and it can be fully charged in 2 hour [22], if we assume a daily commute of 100km. Namely, we have $c = 24$ and $T = 180$.

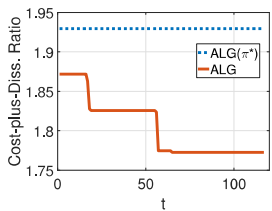
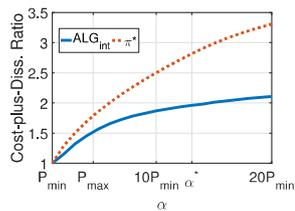
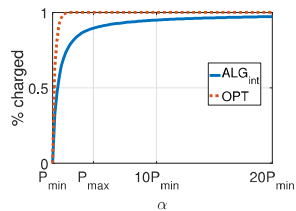
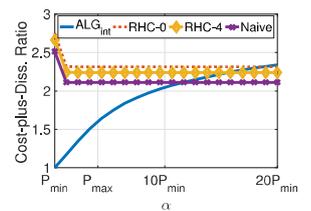

Figure 5: Empirical Cost-plus-Diss. ratio of $ALG(\pi^*)$ and $ALG$ in different slots.

Figure 6: Empirical and theoretical Cost-plus-Diss. ratio of $ALG_{int}$.

Figure 7: Percentage of charged of $ALG_{int}$ and $OPT$ with varying $\alpha$.

Figure 8: Averaged Cost-plus-Diss. ratio of different algorithms with varying $\alpha$.

### B. Evaluation of Online Algorithms

**$ALG(\pi^*)$ VS. $ALG$.** Analysis in Sec. V-A shows that $ALG(\pi^*)$ and $ALG$ has the same competitive ratio. Meanwhile, $ALG$ performs better than $ALG(\pi^*)$ under general-case inputs. To compare their performances, we relax the charging limit constraint in $EVC$ and run the simulation using the real-time price data. We plot the empirical online-to-offline cost-plus-dissatisfaction (denoted as Cost-plus-Diss.) ratio at different slots in Fig. 5. From this figure, we have the following two observations: i) the empirical cost-plus-dissatisfaction ratio of $ALG(\pi^*)$ is fixed over time. This is reasonable because by design, $ALG(\pi^*)$ will try to keep the ratio to be $\pi^*$ at each slot; ii) the empirical cost-plus-dissatisfaction ratio of $ALG$ is non-increasing in $t$ and it is always smaller than that of $ALG(\pi^*)$. This is because $ALG$ can adaptively change the target ratio $\pi_t$ and $\pi_t$ is non-increasing in $t$. These observations verify our results in Sec. V-A and demonstrate the superiority of algorithm $ALG$ over $ALG(\pi^*)$.

**Cost-minimization VS. User Dissatisfaction.** The EV owner can adjust $\alpha$ to strike a balance between cost minimization and dissatisfaction. On the one hand, if $\alpha$ is small, the EV owner prefers to charge the battery at a low cost and can bear the risk of not being fully charged for the EV. On the other hand, if $\alpha$ is large, the EV owner prefers to charge the EV more, despite of potentially high prices and consequently large charging cost. We conduct simulations to understand the impact of $\alpha$.

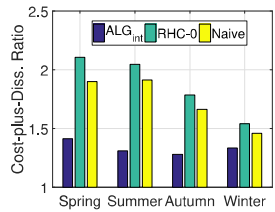
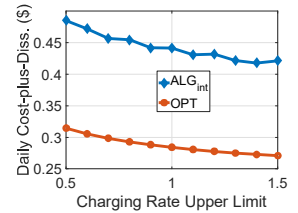

Figure 9: Cost-plus-Diss. Ratio of $ALG_{int}$, $RHC$-0 and $Naive$ online algorithm.

Figure 10: Averaged daily Cost-plus-Diss. of $ALG_{int}$ and $OPT$.

Let us first see how $\alpha$ can affect the performance of $ALG_{int}$, as suggested by our analysis in Sec. V-A. In particular, we vary $\alpha$ in $[p_{\min}, 20p_{\min}]$ and plot the averaged empirical cost-plus-dissatisfaction ratio of $ALG_{int}$ (where each sub-problem runs $ALG$) and its theoretical competitive ratio $\pi^*$ in one year in Fig. 6. We can observe that the empirical cost-plus-dissatisfaction ratio of $ALG_{int}$ is much smaller than $\pi^*$, as in practice the worst case input seldom happens. Meanwhile, both curves grow sublinearly as $\alpha$ increases, i.e., large $\alpha$ degrades the empirical online-to-offline performance ratio for $ALG_{int}$. These observations are aligned with our theoretical results in Theorem 4 and 12. Note that the online-to-offline performance ratio can be interpreted as the performance improvement if future prices are given in advance. Thus this result also highlights that knowing future prices can lead to bigger performance improvement in the case of not considering user dissatisfaction ($\alpha$ is large) than otherwise.

We then investigate the impact of $\alpha$ on the charged percentage, i.e., $\sum_{t=1}^{T} v(t)/c$, at the end of the charging period. Intuitively, we would expect a large (resp. small) charged percentage if $\alpha$ is large (resp. small). Indeed this is the case, as we can see from Fig. 7, which shows the charged percentage (% charged) of $ALG_{int}$ and optimal offline algorithm $OPT$. The charged percentage of $ALG_{int}$ grows as $\alpha$ increases. In particular, when $\alpha = 10p_{\min}$, $ALG_{int}$ charges around 95% of the requested capacity $c$.

**Comparison with Alternatives.** We compare the performance of $ALG_{int}$ with other conceivable alternatives. One is the Receding Horizon Control (RHC) [23] with look-ahead window size $n \in \mathbb{Z}^+$ (denoted as $RHC$-$n$), i.e., at time $t$, given the prices from slots $t$ to $t + n$, $RHC$-$n$ optimizes over these slots and executes only the action at the current slot. In particular, $RHC$-0 in our scenario corresponds to the online strategy of charging the EV at the maximum rate in each slot, regardless of the price. Thus *RHC*-0 can be regarded as our online algorithm $ALG_{int}$ with $\alpha$ being infinitely large. Besides RHC, we also compare our algorithm with a naive threshold-based online algorithm, i.e., the EV owner sets $\frac{p_{max}+p_{min}}{2}$ as the threshold and charges the EV whenever the price is lower than the threshold. We plot the empirical cost-plus-dissatisfaction ratios of different algorithms in Fig. 8. From this figure, we can observe that when $\alpha$ is small, $ALG_{int}$ outperforms other algorithms. However, large $\alpha$ degrades the performance of $ALG_{int}$. Intuitively, this is because in our model, we assume $T$ is unknown. Thus $ALG_{int}$ will try to reserve some capacity for the future good prices. When $\alpha$ is large, $ALG_{int}$ will then suffer a large user dissatisfaction as the EV is not fully charged. In Fig. 9, we fix $\alpha = p_{max}$, and plot the empirical cost-plus-dissatisfaction ratios in different seasons. We can observe that our algorithm achieves substantial performance gain as compared to the alternatives in all the seasons. Moreover, we note that the online-to-offline performance ratio of $ALG_{int}$ is upper-bounded by $\pi^*$, which is better than that of RHC and the naive algorithm.

**Impact of Charging Rate Limit.** In practice, an EV owner may face the question of whether to have a fast charger (large charging power) or a slow charger (small charging power). To address this question, we vary the charging rate upper limit in our simulation and study the average daily cost-plus-dissatisfaction value under different charging power. In particular, we very the normalized charging rate limit from 0.5 to 1.5 (i.e., the charging power of the charger varies from 4.4kW to 13.2kW) and the results are shown in Fig. 10. We can observe that with fast charger, both optimal offline and $ALG_{int}$ have a lower daily cost-plus-dissatisfaction value on average. Meanwhile, increasing EV charging rate limit decreases overall cost almost linearly. Intuitively, this is because a fast charger provides more flexibility, i.e., the algorithm can charge the EV more (resp. less) when the price is low (resp. high), incurring a lower cost-plus-dissatisfaction value as compared to the slow charger case.

## VII. CONCLUSION

In this paper, we investigate the online EV charging problem under real-time pricing, where neither accurate prediction nor distribution of future real-time prices is available to users when making online charging decision. We take into consideration the user dissatisfaction and propose two optimal deterministic online algorithms. The theoretical competitive ratio of our proposed online algorithms is upper bounded by $\min\left\{\sqrt{\alpha/p_{min}}, p_{max}/p_{min}\right\}$, and their empirical online-to-offline performance ratio are shown to outperform other conceivable alternatives, through extensive simulations based on real-world traces.

In terms of future work, there are several directions to explore. First, solving the problem with known $T$ is an interesting direction. Second, it would be interesting to develop online algorithms with minimum commitment, i.e., the online algorithm can guarantee a certain charged percentage at the end of charging period.

## VIII. APPENDIX

### A. Proof of Lemma 2

*Proof:* Since $ALG(\pi)$ outputs 0 whenever $p(t) \geq \alpha$, it is sufficient to consider the input $\sigma^{[1:T]}$ that satisfies $p(\tau) < \alpha, \forall \tau \in [T]$. Given $\sigma^{[1:T]}$, we have $\tilde{OPT}(t) = p_{\min}^t c$. For ease of presentation, given any $\pi$, define $t_1 \triangleq \min\{t | \tilde{OPT}(t) < \frac{\alpha c}{\pi}\}$. From (7), we know that the output of $ALG(\pi)$, denoted as $v(\tau), \tau \in [T]$, can be expressed as

$$v(\tau) = \begin{cases} 0 & , \forall \tau < t_1 \\ \frac{\eta^{\tau-1} - \tilde{OPT}(\tau)\pi}{\alpha - p(\tau)} & , \forall \tau \geq t_1 \end{cases}, \quad (17)$$

and we have $\eta^{t_1-1} = \alpha c$ and $\eta^\tau = \tilde{OPT}(\tau)\pi, \forall \tau \geq t_1$. Namely, for any $\pi$, when $\tau \leq t_1$, the output of $ALG(\pi)$ is zero until $t = t_1$. Thus the summation of outputs of $ALG(\pi)$ in $\sigma^{[1:T]}$ is equivalent to that in $\sigma^{[1:T]}$. Denote the first price in $\sigma^{[t_1:T]}$ as $p_1$, we have $p_1 = p(t_1) \leq \min\{\frac{\alpha}{\pi}, p_{\max}\}$. When $\tau > t_1$, from (17) and (4), we know that the output of $ALG(\pi)$ is positive only when the current price is the lowest price that appear so far. Thus we can delete the slots when the output of $ALG(\pi)$ is zero and the resulting price sequence is decreasing. ∎

### B. Proof of Lemma 3

*Proof:* From (7), under the input $\bar{\sigma}^{[1:T]}$, we know that

$$v(1) = \frac{\alpha c - p(1)c\pi}{\alpha - p(1)},$$

and for all $\tau \in [2, T]$,

$$v(\tau) = \frac{p(\tau-1) - p(\tau)}{\alpha - p(\tau)} c\pi$$
$$= (1 - \frac{\alpha - p(\tau-1)}{\alpha - p(\tau)})c\pi.$$

Then we have

$$V(\pi) = \max_{\bar{\sigma}^{[1:T]}} \sum_{\tau=1}^{T} v(\tau)$$
$$= \max_{\bar{\sigma}^{[1:T]}} \frac{\alpha c - p(1)c\pi}{\alpha - p(1)} + \sum_{\tau=2}^{T}(1 - \frac{\alpha - p(\tau-1)}{\alpha - p(\tau)})c\pi$$

Firstly, note that for any $\tau \in [2, T]$, we have

$$1 - \frac{\alpha - p(\tau-1)}{\alpha - p(\tau)} \leq \ln \frac{\alpha - p(\tau)}{\alpha - p(\tau-1)}.$$

Then we know that given $p(1)$ and $p(T)$, we have

$$\max_{p(2),p(3),\ldots,p(T-1)} \sum_{\tau=2}^{T}(1 - \frac{\alpha - p(\tau-1)}{\alpha - p(\tau)}) = \ln \frac{\alpha - p(T)}{\alpha - p(1)}.$$

Thus we can re-express $V(\pi)$ as the following

$$V(\pi) = \max_{p(1),p(T)} \frac{\alpha c - p(1)c\pi}{\alpha - p(1)} + c\pi \ln \frac{\alpha - p(T)}{\alpha - p(1)}$$
$$= \max_{p(1)} \frac{\alpha c - p(1)c\pi}{\alpha - p(1)} + c\pi \ln \frac{\alpha - p_{\min}}{\alpha - p(1)}.$$

Secondly, note that $p(1) \leq \min\{\frac{\alpha}{\pi}, p_{\max}\}$. Then $V(\pi)$ can be expressed as

$$V(\pi) = \max_{x \leq \min\{\frac{\alpha}{\pi}, p_{\max}\}} f(x)$$
$$= \max_{x \leq \min\{\frac{\alpha}{\pi}, p_{\max}\}} \frac{\alpha c - xc\pi}{\alpha - x} + c\pi \ln \frac{\alpha - p_{\min}}{\alpha - x}.$$

Taking the derivative w.r.t. $x$, we get

$$f'(x) = \frac{\alpha - \pi x}{(\alpha - x)^2}.$$

It is easy to check that $f(x)$ obtain the maximum at $x = \frac{\alpha}{\pi}$. Then we have

$$V(\pi) = \begin{cases} c\pi \ln \frac{\alpha - p_{\min}}{\alpha - \frac{\alpha}{\pi}} & , \frac{\alpha}{\pi} \leq p_{\max} \\ \frac{\alpha c - p_{\max} c\pi}{\alpha - p_{\max}} + c\pi \ln \frac{\alpha - p_{\min}}{\alpha - p_{\max}} & , \frac{\alpha}{\pi} > p_{\max} \end{cases}.$$

It then remains to prove that $V(\pi)$ is decreasing in $\pi$. When $\frac{\alpha}{\pi} \leq p_{\max}$, we have

$$V'(\pi) = c \ln \frac{\alpha - p_{\min}}{\alpha - \frac{\alpha}{\pi}} - c\frac{1}{\pi - 1}$$
$$= c(\ln \frac{\alpha - p_{\min}}{\alpha} - (\ln(1 - \frac{1}{\pi}) + \frac{1}{\pi - 1}))$$

Since $\ln \frac{1}{z} \geq 1 - z$ for any $z \geq 0$, we have

$$\ln(1 - \frac{1}{\pi}) \geq 1 - \frac{\pi}{\pi - 1} = \frac{-1}{\pi - 1}. \quad (18)$$

Thus

$$\ln(1 - \frac{1}{\pi}) + \frac{1}{\pi - 1} \geq 0.$$

It then follows that $V'(\pi) < 0$, thus $V(\pi)$ is decreasing in $\pi$ when $\frac{\alpha}{\pi} \leq p_{\max}$. When $\frac{\alpha}{\pi} > p_{\max}$, we have

$$V'(\pi) = -\frac{p_{\max} c}{\alpha - p_{\max}} + \ln \frac{\alpha - p_{\min}}{\alpha - p_{\max}}$$
$$\leq \frac{p_{\max}(1 - c) - p_{\min}}{\alpha - p_{\max}} < 0.$$

Thus we know that $V(\pi)$ is also decreasing in $\pi$ when $\frac{\alpha}{\pi} > p_{\max}$. ∎

### C. Proof of Theorem 4

*Proof:* We know that $\pi^*$ is the unique solution to the equation $V(\pi^*) = c$. Further, we have

$$V(\frac{\alpha}{p_{\max}}) = \frac{\alpha}{p_{\max}} c \ln \frac{\alpha - p_{\min}}{\alpha - p_{\max}}.$$

It is easy to check that $V(\frac{\alpha}{p_{\max}})$ is decreasing in $\alpha$. From the monotonicity of $V(\pi)$, we know that if $V(\frac{\alpha}{p_{\max}}) < c$, then $\alpha > \pi p_{\max}$, otherwise we have $\alpha \leq \pi p_{\max}$. Denote $\alpha^*$ as the root for equation $\frac{\alpha}{p_{\max}} \ln \frac{\alpha - p_{\min}}{\alpha - p_{\max}} = 1$, then we have $V(\frac{\alpha^*}{p_{\max}}) = c$. Now when $\alpha > \alpha^*$, we know $V(\frac{\alpha}{p_{\max}}) < V(\frac{\alpha^*}{p_{\max}}) = c$, then $\pi^*$ should satisfy

$$V(\pi^*) = \frac{\alpha - p_{\max} \pi^*}{\alpha - p_{\max}} c + c\pi^* \ln \frac{\alpha - p_{\min}}{\alpha - p_{\max}} = c.$$

Solving the linear equation we get

$$\pi^* = \frac{\frac{p_{\max}}{\alpha - p_{\max}}}{\frac{p_{\max}}{\alpha - p_{\max}} - \ln \frac{\alpha - p_{\min}}{\alpha - p_{\max}}}.$$

When $\alpha \leq \alpha^*$, we have

$$V(\pi^*) = c\pi^* ln\frac{\alpha - p_{\min}}{\alpha - \frac{\alpha}{\pi^*}} = c,$$

thus $\pi^*$ is the root for the equation $\pi^* ln\frac{\alpha - p_{\min}}{\alpha - \frac{\alpha}{\pi^*}} = 1$.

It then remains to show that $\pi^*$ is upper bounded by $\min\{\theta, \sqrt{\frac{\alpha}{p_{\min}}}\}$. Firstly, when $\alpha > \alpha^*$, we have $\pi^* < \frac{\alpha}{p_{\max}}$. One can easily check that $\pi^*$ defined in (12) is increasing in $\alpha$. Further, we have

$$\lim_{\alpha \to \infty} \pi^* = \lim_{\alpha \to \infty} \frac{1}{1 - \frac{\alpha - p_{\max}}{p_{\max}} \cdot (1 - \frac{\alpha - p_{\max}}{\alpha - p_{\min}})}$$
$$= \lim_{\alpha \to \infty} \frac{1}{1 - \frac{\alpha - p_{\max}}{\alpha - p_{\min}} \cdot \frac{p_{\max} - p_{\min}}{p_{\max}}}$$
$$= \frac{p_{\max}}{p_{\min}} = \theta.$$

Namely, when $\alpha > \alpha^*$, $\pi^*$ is upper bounded by $\theta$. In particular, when $\alpha \in (\alpha^*, p_{\max}\theta)$, we can have a tighter bound. Recall that when $\alpha > \alpha^*$, we know $\pi^* < \frac{\alpha}{p_{\max}}$, it then follows that

$$\pi^* < \frac{\alpha}{p_{\max}} < \theta = \frac{p_{\max}}{\alpha} \frac{\alpha}{p_{\min}} < \frac{\alpha}{p_{\min}\pi^*}.$$

So we have $\pi^* \leq \sqrt{\frac{\alpha}{p_{\min}}} \leq \theta$, when $\alpha \in (\alpha^*, p_{\max}\theta)$. When $\alpha \geq p_{\max}\theta$, since $\sqrt{\frac{\alpha}{p_{\min}}} \geq \theta$, we know $\pi^* \leq \theta$ is a tighter bound. In conclusion, when $\alpha > \alpha^*$, $\pi^*$ is upper bounded by $\min\{\theta, \sqrt{\frac{\alpha}{p_{\min}}}\}$.

Secondly, when $\alpha \leq \alpha^*$, we know $\ln\frac{\alpha - p_{\min}}{\alpha - \frac{\alpha}{\pi^*}} = \frac{1}{\pi^*}$. Taking the exponential on both sides, we have

$$\frac{\alpha - p_{\min}}{\alpha - \frac{\alpha}{\pi^*}} = e^{\frac{1}{\pi^*}} \geq \frac{1}{\pi^*} + 1,$$

where the inequality follows from the fact that $e^x \geq x+1, \forall x \in \mathbb{R}$. Solving the above inequality, we get $\pi^* \leq \sqrt{\frac{\alpha}{p_{\min}}}$. This completes our proof. ∎

### D. Proof of Theorem 5

*Proof:* Consider an arbitrary deterministic online algorithm different from $ALG(\pi^*)$, denoted as $\mathcal{A}$. Using an adversary argument we show that $\mathcal{A}$ cannot achieve a ratio smaller than $\pi^*$. For $\mathcal{A}$ and $ALG(\pi^*)$, denote the output at time $t$ as $v_{\mathcal{A}}(t)$ and $v(t)$, respectively. For ease of presentation, denote $\tilde{\sigma}^{[1:T]} = (\tilde{p}(1), \tilde{p}(2), ..., \tilde{p}(T))$ as the worst case input for $ALG(\pi^*)$, i.e., under this input, we have $\sum_{\tau=1}^{T} v(\tau) = c$. According to Lemma 2, we must have $\min\{\frac{\alpha}{\pi}, p_{\max}\} \geq \tilde{p}(1) > \tilde{p}(2) > \cdots > \tilde{p}(T) \geq p_{\min}$.

We present $\tilde{p}(1)$ to $\mathcal{A}$ at the first slot. If $v_{\mathcal{A}}(1) \leq v(1)$, then we end the charging period, i.e., $T = 1$. In this case, the competitive ratio of $\mathcal{A}$ is at least $\pi^*$. Otherwise, if $v_{\mathcal{A}}(1) > v(1)$, we continue to present $\tilde{p}(2)$ to $\mathcal{A}$. In general, if at time $t$ the total amount of energy $\mathcal{A}$ charged is no larger than $\sum_{\tau=1}^{t} v(\tau)$, we immediately end the charging period. Otherwise, we continue and present $\mathcal{A}$ with the next price. Let $t'$ be the minimum $t$ such that at the end of time $t$, the total amount of energy $\mathcal{A}$ charged is less than $\sum_{\tau=1}^{t} v(\tau)$. We note that $t'$ must exist, otherwise we have at time $T$, $\sum_{\tau=1}^{T} v_{\mathcal{A}}(\tau) > \sum_{\tau=1}^{T} v(\tau) = c$ is a contradiction. Then we have

$$v_{\mathcal{A}}(1) > v(1)$$
$$\sum_{\tau=1}^{2} v_{\mathcal{A}}(\tau) > \sum_{\tau=1}^{2} v(\tau)$$
$$\cdots$$
$$\sum_{\tau=1}^{t'-1} v_{\mathcal{A}}(\tau) > \sum_{\tau=1}^{t'-1} v(\tau)$$

and by the definition of $t'$, we have

$$\sum_{\tau=1}^{t'} v_{\mathcal{A}}(\tau) \leq \sum_{\tau=1}^{t'} v(\tau). \quad (19)$$

Since $\tilde{p}(\tau)$ are decreasing in $\tau$, $\mathcal{A}$ would have achieved a lower cost-plus-dissatisfaction by charging exactly $v(\tau)$ for any $\tau \in [t'-1]$ and by charging $v^*_{\mathcal{A}}(t') = v_{\mathcal{A}}(t') + \sum_{\tau=1}^{t'-1} v_{\mathcal{A}}(\tau) - \sum_{\tau=1}^{t'-1} v(\tau)$ at time $t'$. Namely, we have

$$\eta^{t'}_{\mathcal{A}} \geq \alpha c - \sum_{\tau=1}^{t'-1}(\alpha - \tilde{p}(\tau))v(\tau) - (\alpha - \tilde{p}(t'))v^*_{\mathcal{A}}(t'),$$

where $\eta^{t'}_{\mathcal{A}}$ is the cost-plus-dissatisfaction of $\mathcal{A}$ at time $t'$. However, from (19), we know $v^*_{\mathcal{A}}(t') \leq v(t')$ and thus we have

$$\eta^{t'}_{\mathcal{A}} \geq \alpha c - \sum_{\tau=1}^{t'}(\alpha - \tilde{p}(\tau))v(\tau) = O\tilde{P}T(t')\pi^*.$$

Thus the competitive ratio of $\mathcal{A}$ should at least be $\pi^*$.

It follows that $\mathcal{A}$ must coincide with $ALG(\pi^*)$, achieving a ratio of $\pi^*$, or otherwise $\mathcal{A}$ incurs a higher ratio on $\tilde{\sigma}^{[1:T]}$. ∎

### E. Proof of Lemma 6

*Proof:* We first show the expression of $V_t(\pi_t)$. Given any $\pi_t$, we know that the output at time $t+1$ needed to maintain the online to offline cost-plus-dissatisfaction ratio to be $\pi_t$ is

$$v(t+1) = \frac{\eta^t - p(t+1)c\pi_t}{\alpha - p(t+1)}$$

and for any $\tau \geq t+2$, we have

$$v(\tau) = \frac{p(\tau-1) - p(\tau)}{\alpha - p(\tau)}c\pi_t \leq c\pi_t \ln\frac{\alpha - p(\tau)}{\alpha - p(\tau-1)}.$$

Then given any $p(t+1)$, we have

$$\max_{\tilde{\sigma}^{[t+2,T]}} \sum_{\tau=t+1}^{T} v(\tau) = \frac{\eta^t - p(t+1)c\pi_t}{\alpha - p(t+1)} + c\pi_t \ln\frac{\alpha - p_{\min}}{\alpha - p(t+1)}.$$

It then follows that

$$\max_{\tilde{\sigma}^{[t+1,T]}} \sum_{\tau=t+1}^{T} v(\tau) = \max_{p(t+1)<p(t)} \frac{\eta^t - p(t+1)c\pi_t}{\alpha - p(t+1)} + c\pi_t \ln\frac{\alpha - p_{\min}}{\alpha - p(t+1)}.$$

Define $h(x) = \frac{\eta^t - xc\pi_t}{\alpha - x} + c\pi_t \ln \frac{\alpha - p_{\min}}{\alpha - x}$. We have $h'(x) = \frac{xc\pi_t - \eta^t}{(\alpha-x)^2}$. Thus $h(x)$ obtain the maximum at $x = \frac{\eta^t}{c\pi_t}$. Note that we have $\eta^t = p(t)c\pi_t$, thus we know that

$$V_t(\pi_t) = \sum_{\tau=1}^{t-1} v(\tau) + v(t) + \max_{\vec{\sigma}^{[t+1,T]}} \sum_{\tau=t+1}^{T} v(\tau)$$

$$= \sum_{\tau=1}^{t-1} v(\tau) + v(t) + c\pi_t \ln \frac{\alpha - p_{\min}}{\alpha - \frac{\eta_t}{c\pi_t}}$$

$$= \sum_{\tau=1}^{t-1} v(\tau) + \frac{\eta^{t-1} - p(t)c\pi_t}{\alpha - p(t)} + c\pi_t \ln \frac{\alpha - p_{\min}}{\alpha - p(t)}$$

We then show that $V_t(\pi_t)$ is decreasing in $\pi_t$. Note that $V_t(\pi_t)$ is a linear function on $\pi_t$ and the first order derivative is

$$c\left(\ln \frac{\alpha - p_{\min}}{\alpha - p(t)} - \frac{p(t)}{\alpha - p(t)}\right).$$

It is easy to check that

$$\frac{p(t)}{\alpha - p(t)} = \frac{\alpha}{\alpha - p(t)} - 1 \geq \ln \frac{\alpha}{\alpha - p(t)} > \ln \frac{\alpha - p_{\min}}{\alpha - p(t)}. \quad (20)$$

It then follows that $V_t(\pi_t)$ is decreasing in $\pi_t$. ∎

*F. Proof of Theorem 7*

*Proof:* Since $V_t(\pi_t)$ is linearly decreasing in $\pi_t$, we have $V_t(\pi_t^*) = c$. Then we can get the expression of $\pi_t^*$ in (16). It then follows that, for any $\pi_{t-1}^*$ and $\pi_t^*$, we have

$$V_{t-1}(\pi_{t-1}^*) = \sum_{\tau}^{t-1} v(\tau) + c\pi_{t-1}^* \ln \frac{\alpha - p_{\min}}{\alpha - p(t-1)} = c,$$

$$V_t(\pi_t^*) = \sum_{\tau}^{t} v(\tau) + c\pi_t^* \ln \frac{\alpha - p_{\min}}{\alpha - p(t)} = c.$$

Subtracting the above two equations, we get

$$0 = v(t) + c\pi_t^* \ln \frac{\alpha - p_{\min}}{\alpha - p(t)} - c\pi_{t-1}^* \ln \frac{\alpha - p_{\min}}{\alpha - p(t-1)}$$

$$= \frac{p(t-1)\pi_{t-1}^* - p(t)\pi_t^*}{\alpha - p(t)} + \pi_t^* \ln \frac{\alpha - p_{\min}}{\alpha - p(t)}$$

$$- \pi_{t-1}^* \ln \frac{\alpha - p_{\min}}{\alpha - p(t-1)}$$

$$= (\pi_t^* - \pi_{t-1}^*) \ln \frac{\alpha - p_{\min}}{\alpha - p(t)} + \frac{p(t-1)\pi_{t-1}^* - p(t)\pi_t^*}{\alpha - p(t)}$$

$$- \pi_{t-1}^* \ln \frac{\alpha - p(t)}{\alpha - p(t-1)}$$

Since $\ln \frac{\alpha - p(t)}{\alpha - p(t-1)} \geq 1 - \frac{\alpha - p(t-1)}{\alpha - p(t)}$, we have

$$(\pi_t^* - \pi_{t-1}^*)(\ln \frac{\alpha - p_{\min}}{\alpha - p(t)} - \frac{p(t)}{\alpha - p(t)}) \geq 0.$$

From (20) we know that $\ln \frac{\alpha - p_{\min}}{\alpha - p(t)} \leq \frac{p(t)}{\alpha - p(t)}$, then we must have for any $t \in [T]$, $\pi_t^* \leq \pi_{t-1}^*$. ∎

*G. Proof of Lemma 9*

*Proof:* As each $EVC_i$ follows $ALG(\pi^*)$ if it's assigned with a price and remains unchanged otherwise, from Corollary 8, we easily conclude it. ∎

*H. Proof of Lemma 10*

*Proof:* Firstly, we claim $v(t) = \sum_{i=1}^{c} v_i(t), \forall t \in [T]$. To see this, consider the following two cases:

Case I, the condition at line 4 in Algorithm 1 fails and $p(t)$ is assigned to a $EVC_i$. In this case, $v(t) = v_i(t)$ and for any other $EVC_i$, $v_i(t) = 0$. So $v(t) = \sum_{i=1}^{c} v_i(t)$.

Case II, the condition at line 4 in Algorithm 1 holds and $p(t)$ is not assigned to any $EVC_i$. In this case, $v(t) = 0$ and for any $EVC_i$, $v_i(t) = 0$. So $v(t) = \sum_{i=1}^{c} v_i(t)$.

Secondly, observe that at time $t$, by definition, we have

$$\eta^t = \sum_{\tau=1}^{t} p(\tau)v(\tau) + \alpha\left(c - \sum_{\tau=1}^{t} v(\tau)\right)$$

$$= \sum_{\tau=1}^{t}\left(p(\tau)\sum_{i=1}^{c} v_i(\tau)\right) + \alpha\left(c - \sum_{\tau=1}^{t}\sum_{i=1}^{c} v_i(\tau)\right)$$

$$= \sum_{i=1}^{c}\left(\sum_{\tau=1}^{t} p(\tau)v_i(\tau) + \alpha\left(1 - \sum_{\tau=1}^{c} v_i(\tau)\right)\right) = \sum_{i=1}^{c} \eta_i^t.$$

This completes our proof. ∎

*I. Proof of Lemma 11*

*Proof:* For ease of presentation, we present the proof for the case that $p(\tau) < \alpha, \forall \tau \in [T]$. The case that there exist $p(\tau) \geq \alpha$ is trivial, since both $OPT(\tau)$ and $OPT_i(\tau), \forall i \in [c]$ remains unchanged at time $\tau$.

Firstly, note that when $p(\tau) < \alpha, \forall \tau \in [T]$, we have $\mathcal{T}_t^\alpha = [t]$ for any $t \in [T]$, thus the set $\mathcal{T}_t \cap \mathcal{T}_t^\alpha = \mathcal{T}_t$. Namely, it is sufficient to consider the set $\mathcal{T}_t$.

Initially, $OPT(0) = \alpha c = \sum_{i=1}^{c} \alpha = \sum_{i=1}^{c} OPT_i(0)$.

When $t \leq c$, we know that $\mathcal{T}_t = [t]$. The offline cost-plus-dissatisfaction can be expressed as

$$OPT(t) = \sum_{\tau=1}^{t} p(\tau) \times 1 + \alpha(c - \sum_{\tau=1}^{t} 1).$$

Meanwhile, following $ALG_{int}$, it's easy to see that at slot $t \leq c$, $i^* = \arg\max_{i \in [c]} \mu_i^{t-1} = t$ as $\alpha > p(\tau), \forall \tau \in [T]$, and thus $p(t)$ is assigned to $EVC_t$. As a result, if $i \leq t$, $OPT_i(t) = p(i)$ as $EVC_i$ is assigned with $p(i)$ only up to time $t$; if $i > t$, $EVC_i$ is not assigned any $p(t)$ yet and $OPT_i(t) = OPT_i(0) = \alpha$ as $EVC_i$ haven't been assigned with any price. Thus we have for any $t \leq c$,

$$\sum_i OPT_i(t) = \sum_{i=1}^{t} p(i) + \alpha \sum_{i=t+1}^{c} 1$$

$$= \sum_{\tau=1}^{t} p(\tau) \times 1 + \alpha(c - \sum_{\tau=1}^{t} 1) = OPT(t).$$

When $t \geq c$, we know that $OPT(t) = \sum_{\tau \in \mathcal{T}_t} p(\tau)$. We also have $OPT_i(t) = \mu_i^t$. It's then sufficient to show that multiset

$$\{\mu_i^t, i \in [c]\} = \{p(\tau), \tau \in \mathcal{T}_t\}. \quad (21)$$

We prove it by mathematics induction. At slot $t = c$, $\mu_i^t = p(i), \forall i \in [c]$ and $\mathcal{T}_t = [c]$ as there are exactly $c$ input prices. So we know (21) holds for $t = c$. Suppose at time $t - 1$, we have $\{\mu_i^{t-1}, i \in [c]\} = \{p(\tau), \tau \in \mathcal{T}_{t-1}\}$. Then at time $t$, consider the following two cases:

Case I: $p(t) \geq \max\{\mu_i^{t-1}, i \in [c]\}$, or equivalently, $p(t) \geq \max\{p(\tau), \tau \in \mathcal{T}_{t-1}\}$. In this case, we know that $\mathcal{T}_t = \mathcal{T}_{t-1}$ and the price $p(t)$ is discarded according to the input distributor in $ALG_{int}$. Thus we have $\mu_i^t = \mu_i^{t-1}$ for all $i \in [c]$. Then we know that (21) holds for $t$ as well in this case.

Case II: $p(t) < \max\{\mu_i^{t-1}, i \in [c]\}$, or equivalently, $p(t) < \max\{p(\tau), \tau \in \mathcal{T}_{t-1}\}$. According to the offline solution, we know that

$$\{p(\tau), \tau \in \mathcal{T}_t\} = \{p(\tau), \tau \in \mathcal{T}_{t-1}\} - \{p(i^*)\} + \{p(t)\}.$$

where $i^* = \arg\max_{\tau \in \mathcal{T}_{t-1}} p(\tau)$. Since (21) holds for $t - 1$, we have $\max_{\tau \in \mathcal{T}_{t-1}} p(\tau) = \max_{i \in [c]} \mu_i^{t-1}$, or equivalently, $p(i^*) = \mu_{i^*}^{t-1}$. It then follows that

$$\begin{aligned}\{p(\tau), \tau \in \mathcal{T}_t\} &= \{\mu_i^{t-1}, i \in [c]\} - \{p(i^*)\} + \{p(t)\} \\ &= \{\mu_i^{t-1}, i \in [c]\} - \{u_{i^*}^{t-1}\} + \{p(t)\} \\ &= \{\mu_i^t, i \in [c]\},\end{aligned}$$

where the last equality follows from the input distributor in $ALG_{int}$ as it distribute $p(t)$ to $EVC_{i^*}$. We conclude that (21) holds for $t$ as well in this case. ∎

*J. Proof of Theorem 13*

*Proof:* The optimality is proved as following. It's similar to the proof of the optimality of $ALG(\pi^*)$ in the non-charging-limit case. Consider the worst case input for $ALG(\pi^*)$ in the non-charging-limit case, denoted as $\tilde{\sigma}^{[1:T]}$. Under this input, we have $V(\pi^*) = c$. Base on $\tilde{\sigma}^{[1:T]}$, we construct another input by repeating each $p(\tau), \tau \in [T]$ in $\tilde{\sigma}^{[1:T]}$ for $c$ times. Denote the new constructed input as $\bar{\sigma}^{[1:cT]}$. It is obvious that under $\bar{\sigma}^{[1:cT]}$, the EV will be fully charged under the online algorithm $ALG_{int}$. Namely, $\bar{\sigma}^{[1:cT]}$ is the worst case input for $ALG_{int}$ in problem $EVC$.

Now consider an arbitrary deterministic online algorithm for $EVC$, denoted as $\mathcal{A}$. Using an adversary argument we show that $\mathcal{A}$ cannot achieve a ratio smaller than $\pi^*$. For $\mathcal{A}$ and $ALG_{int}$, denote the output at time $t$ as $v_{\mathcal{A}}(t)$ and $v(t)$, respectively. For ease of presentation, we define

$$v'_{\mathcal{A}}(i) = \sum_{t=(i-1)c+1}^{ic} v_{\mathcal{A}}(t),$$

$$v'(i) = \sum_{t=(i-1)c+1}^{ic} v(t).$$

Namely, $v'_{\mathcal{A}}(i)$ and $v'(i)$ are the accumulate output from $t = (i-1)c+1$ to $ic$ for $\mathcal{A}$ and $ALG_{int}$, respectively. We say $[(i-1)c+1, ic]$ as period $i$, obviously, $i \in [T]$. We first present the input $\bar{\sigma}^{[1:c]}$ to $\mathcal{A}$. If $v'_{\mathcal{A}}(1) < v'(1)$, then we end the charging period, i.e., $T = 1 \times c$. In this case, the competitive ratio of $\mathcal{A}$ is at least $\pi^*$. Otherwise, if $v'_{\mathcal{A}}(1) \geq v'(1)$, we continue to present $\bar{\sigma}^{[c+1,2c]}$ to $\mathcal{A}$. In general, if at time $i \times c$ the total amount of energy $\mathcal{A}$ charged is less than $\sum_{\tau=1}^{i} v'(\tau)$, we immediately end the charging period. Otherwise, we continue and present $\mathcal{A}$ with the sequence of prices in next period, i.e., $\bar{\sigma}^{[ic+1,(i+1)c]}$. Let $t'$ be the minimum period $t$ such that at the end of time $t' \times c$, the total amount of energy $\mathcal{A}$ charged is less than $\sum_{\tau=1}^{t'} v'(\tau)$.

It is easy to check by contradiction that period $t'$ must exist. Then we have

$$v'_{\mathcal{A}}(1) > v'(1)$$

$$\sum_{\tau=1}^{2} v'_{\mathcal{A}}(\tau) > \sum_{\tau=1}^{2} v'(\tau)$$

$$\cdots$$

$$\sum_{\tau=1}^{t'-1} v'_{\mathcal{A}}(\tau) > \sum_{\tau=1}^{t'-1} v'(\tau)$$

and by the definition of $t'$, we have

$$\sum_{\tau=1}^{t'} v'_{\mathcal{A}}(\tau) \leq \sum_{\tau=1}^{t'} v'(\tau). \tag{22}$$

Since $\bar{p}(\tau)$ are non-increasing in $\tau$, $\mathcal{A}$ would have achieved a lower cost-plus-dissatisfaction by charging exactly $v'(\tau)$ for any $\tau \in [t'-1]$ and by charging $v^*_{\mathcal{A}}(t') = v'_{\mathcal{A}}(t') + \sum_{\tau=1}^{t'-1} v'_{\mathcal{A}}(\tau) - \sum_{\tau=1}^{t'-1} v'(\tau)$ for period $t'$. Namely, we have

$$\eta^{t'}_{\mathcal{A}} \geq \alpha c - \sum_{\tau=1}^{t'-1}(\alpha - \bar{p}(\tau c))v'(\tau) - (\alpha - \bar{p}(t'c))v^*_{\mathcal{A}}(t'),$$

where $\eta^{t'}_{\mathcal{A}}$ is the cost of $\mathcal{A}$ at the end of period $t'$. However, from (22), we have

$$v^*_{\mathcal{A}}(t') \leq v(t')$$

and thus we have

$$\eta^{t'}_{\mathcal{A}} \geq \alpha c - \sum_{\tau=1}^{t'}(\alpha - \bar{p}(\tau c))v(\tau) = OPT(t'c)\pi^*.$$

Thus the competitive ratio of $\mathcal{A}$ should at least be $\pi^*$.

It follows then that $\mathcal{A}$ must coincide with $ALG_{int}$, achieving a ratio of $\pi^*$, or otherwise $\mathcal{A}$ incurs a higher ratio on input $\bar{\sigma}^{[1:cT]}$. ∎